\documentclass[final,3p,times,twocolumn]{elsarticle}

\usepackage{amssymb}
\usepackage{graphicx}
\usepackage{epsfig}

\def\GvC     {{\mbox{$\mathrm{GeV/c}$}}}

\journal{NIM-A}

\begin{document}

\begin{frontmatter}



\title{Measuring the masses of the charged hadrons using a RICH as a precision
       velocity spectrometer}       


\author{Peter.~S.~Cooper}
\address{Fermi National Accelerator Laboratory, Batavia, IL 60510, U.S.A.}

\author{J\"urgen~Engelfried}
\address{Universidad Aut\'onoma de San Luis Potos\'{\i}, San Luis Potos\'{\i}, Mexico}

\begin{abstract}
The Selex experiment measured several billion charged hadron tracks with a high
precision magnetic momentum spectrometer and high precision RICH velocity 
spectrometer.  We have analyzed these data to simultaneously measure the 
masses of all the long lived charged hadrons and anti-hadrons from the $\pi$ 
to the $\Omega$ using the same detector and technique. The statistical 
precision achievable with this data sample is more than adequate for $0.1\%$
mass measurements.

We have used these measurements to develop and understand the systematic 
effects in using a RICH as a precision velocity spectrometer with the goal of 
measuring 10 masses with precision ranging from 100 KeV for the lightest 
to 1000 KeV for the heaviest. This requires controlling the radius measurement 
of RICH rings to the $\sim10^{-4}$ level. Progress in the mass 
measurements and the required RICH analysis techniques developed are discussed.
\end{abstract}

\end{frontmatter}



\begin{figure}[!htb]
\centerline{\psfig{figure=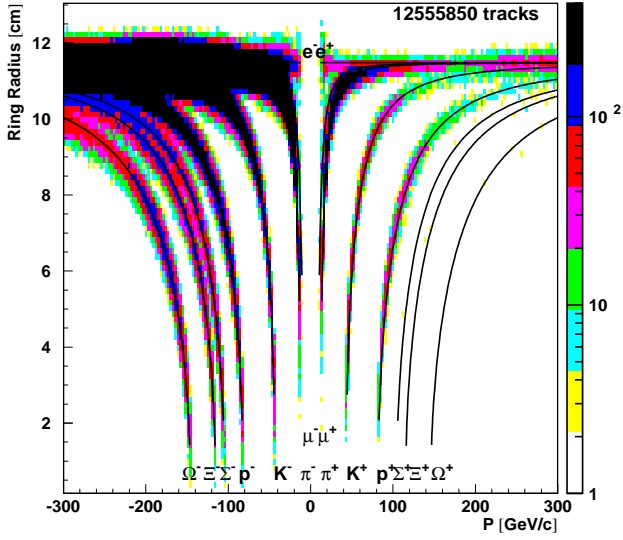,width=3.2in,height=3.2in}}
\caption{RICH Ring radius vs measured momentum for $12.5$M tracks.}
\label{PART3}
\end{figure}

%
     The Selex RICH was orginally conceived as a particle identifier for a 
fixed target multi-particle spectrometer~\cite{Engelfried:1997rp}.  Once we 
saw real data from this detector we learned that this technique had serious 
potential as a precision velocity spectrometer~\cite{Engelfried:2002mh,
Engelfried:2002xh,Cooper:2004hz,Morelos:2005ey}.
Plots like Figure~\ref{PART3}~\cite{Engelfried:2002mh} clearly demonstrated
that precision mass measurements of many of the hadrons, at the same time with 
the same detectors, was possible.  This paper is an exploration of the  
systematic limits of this technique.
 
    Recently the MIPP experiment proposed further data taking to resolve the 
$\sim100$ ppm discrepancy in the latest charged kaon x-ray mass 
measurements~\cite{MIPP}.  They now have the Selex RICH~\cite{SELEX} (with 
$CO_{2}$ instead of $Ne$ as a radiator) but with insufficient data to reach 
the $100$ ppm level of statistical precision.  Selex has large amounts of data 
already recorded, and analysed, in a well understood apparatus.  Selex events \
are multi-hadronic interactions, not single tracks.  Nonetheless we thought it 
would be good to see how far we could take the idea of making precision 
particle mass measurements using a precision momentum (magnetic) and velocity 
(RICH) spectrometers.  We have more than enough data to make mass measurements 
with statistical precsion better than the $<1000$ppm$ = 0.1\%$ level, even 
with very tight cuts to select clean individual track measurements.

     The goals of this study are to identify the important systematic 
uncertainties of RICHs used a precision velocity spectrometers.  Only studies
with real data can fully illuminate the the systematic limitations of such 
precision spectrometry.  It worth noting that $m=p/(\beta\gamma)$, where $p$ 
is the momentum, and $\beta\gamma$ the relativisitic velocity, so that 
systematics in the magnetic momentum spectrometry are equally important: both
must be in control to achieve a precision mass measurement to the $<100$ ppm 
level.

%

   Selex was a Fermilab fixed target experiment designed to study the 
production and decay of charmed baryons.  It took data in $1996-7$ in a 
$600~\GvC$ $\Sigma^{-}$ beam with excellent vertex and momentum 
spectrometers.  The Selex RICH, one of the first large multi-pixel PMT RICHes,
provided $\sim1\%$ velocity resolution for all particles above threshold 
($\beta\gamma>86$).  This is well matched to the $0.5-1\%$ momentum 
resolution for tracks which reach the RICH.   

     The same detectors and analysis provide common systematics for all
particles species.  Figure~\ref{PART3} displays 18 particles species from a
12.5M track sample with RICH Ring Radius plotted as a function of measured 
track momentum.  We have no sensitivity to the electron and muon masses and we 
can't resolve the $\Sigma^{+}$ from the anti-$\Sigma^{-}$.  We can measure the
masses of the other 10 particles.

%

The usual small angle and ultra-relativistic approximations 
[$\Theta_{c}=\sqrt{2\delta-1/\gamma^{2}}$] to the Cherenkov equation 
[$cos(\Theta_{c})=1/\beta n$ $\delta=n-1=67x10^{-6}$] is good to $0.05\%$.
The measured ring radius is related through the spherical mirror focal length
($F$) to the momentum, mass and maximum Cherenkov angle by
$R(p) = F\sqrt{ (\Theta_{c}^{max})^{2} - (m/p)^{2} }$.

Exploiting this relationship to measure mass requires three calibrations; the 
momentum scale, determined by reconstructing $K^{0}_{s}$ decays, the mirror 
focal length, which varies from $989-992cm$ across the 16 mirrors in the Selex
RICH, and the maximum Cherenkov angle $\Theta_{c}^{max}$.

\begin{figure}[!htb]
\begin{minipage}[t]{3.2in}
\psfig{figure=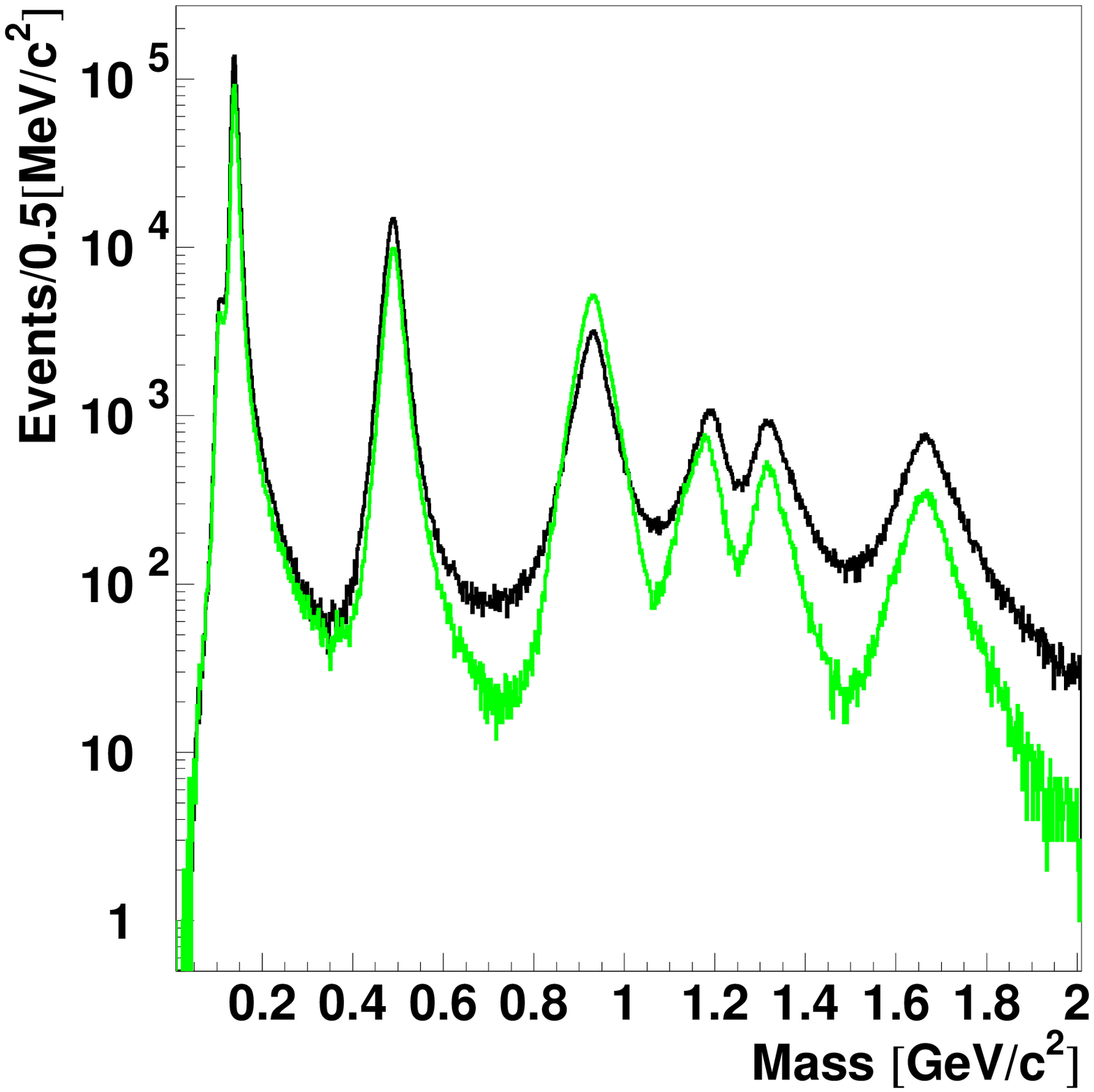,width=3.2in,height=2.5in}
\end{minipage}
\hfill
\begin{minipage}[t]{3.2in}
\psfig{figure=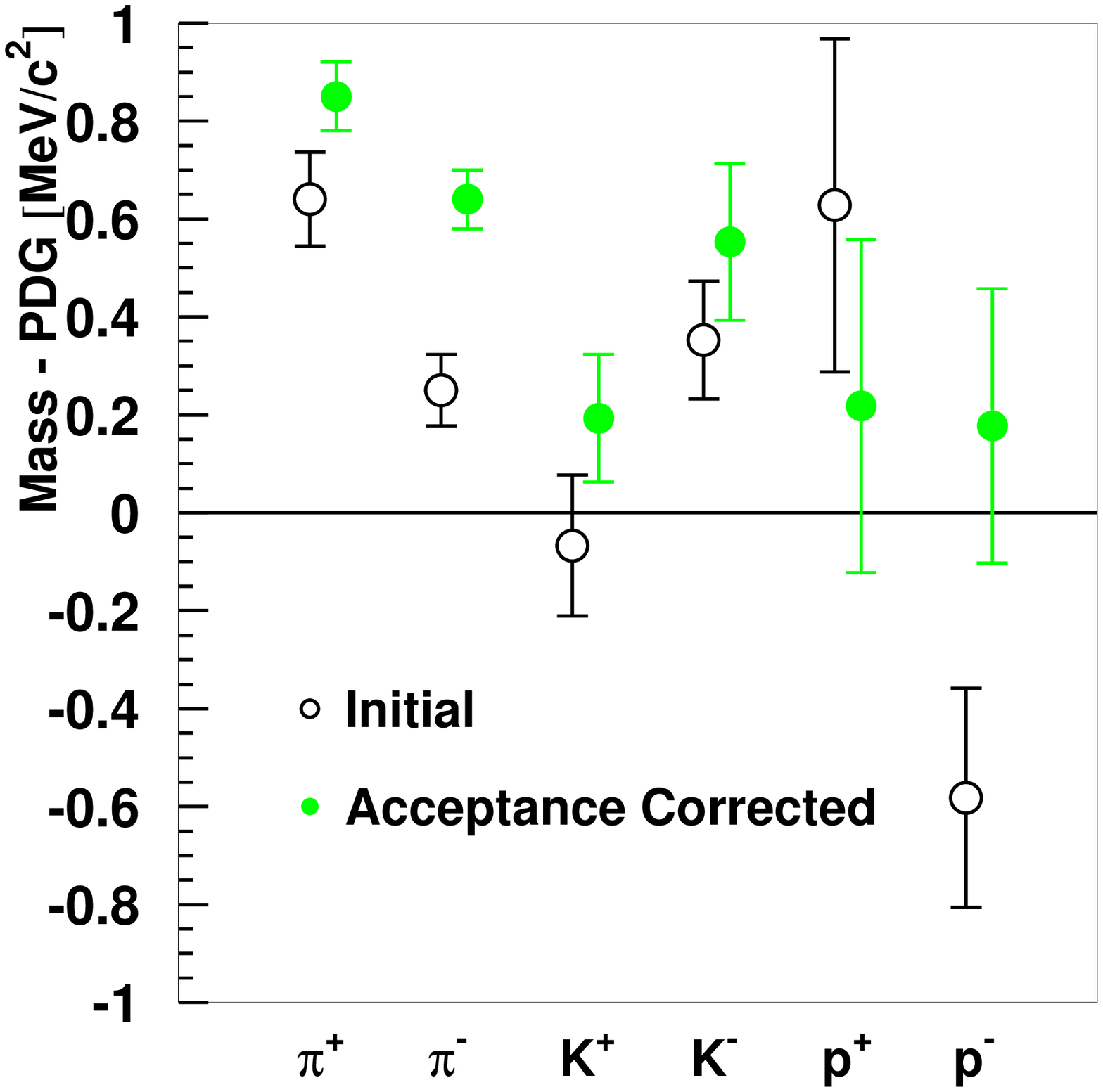,width=3.2in,height=2.5in}
\end{minipage}
\caption{Mass from measured ring radius and momentum [top].
         Fits to masses in 2mm radius bins; open (before), and  solid (after)
         acceptance corrections [bottom].}
\label{MASS}
\end{figure}

Rings are fit for the radius and center coordinates~\cite{RFIT} from a PMT 
list generated around the measured track angles in the RICH cut around the 
predicted radius for a given mass hypothesis.  At least 5 hits are required.  
As a clean place to start the Selex data used are low statistics, low rate, 
low multiplicity data with one of the two spectrometer magnets off.  A mass 
spectrum generated after an initial calibration for tracks with ring radii in 
the interval $5-8 cm$ ($R^{max}=11.4 cm$) is shown in Figure~\ref{MASS}(top).
Gaussian mass fits in $2 mm$ R bins give poor $\chi^{2}$ mass averages shown 
in Figure~\ref{MASS}(bottom, open black points). The difference with the 
PDG~\cite{PDG} mass values are plotted.  The statistical mass uncertainties 
are ($\pi=70~KeV/c^2$, $K=160~KeV/c^2$ and $p=350~KeV/c^2$).  Systematic 
uncertainties clearly dominate these mass measurements.

\begin{figure}[!htb]
\centerline{\psfig{figure=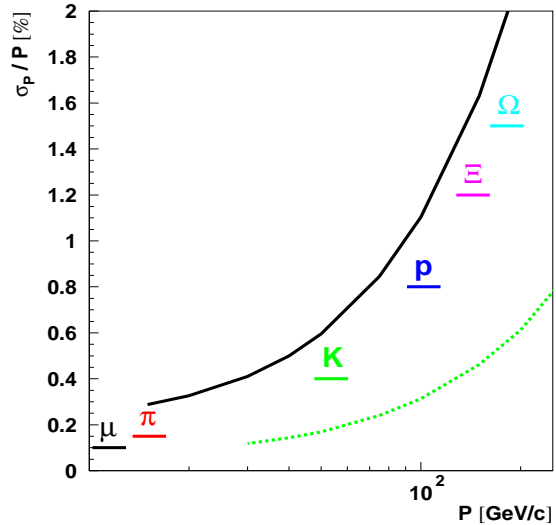,width=3.2in,height=3.0in}}
\caption{Momentum resolution for the data set analysed here (solid) and
        for the Selex charm data (dashed).  The momentum region corresponding 
        to $5<R(cm)<8$ for each particle species is also shown.}
\label{P_RES}
\end{figure}

The momentum resolution is shown in Figure~\ref{P_RES}.  For the data set 
analysed here the momentum cutoff for a track to reach the RICH is $12~GeV/c$.
For the Selex charm data the resolution is much better but the cutoff is 
$20~GeV/c$, putting the pions out of reach but greatly improving the momentum 
resolution for the hyperons.  The momentum scale was calibrated using 
$K_{s}^{0} \rightarrow \pi^{+}\pi^{-}$ decays and the $K_{s}^{0}$ mass from 
the PDG~\cite{PDG}.  With this calibration the masses reconstructed with 
charged particle decays of 12 other hadrons from the $\phi$ through the 
$\Omega_{c}^0$ are correctly reproduced.

\begin{figure}[!htb]
\begin{minipage}[t]{3.2in}
\psfig{figure=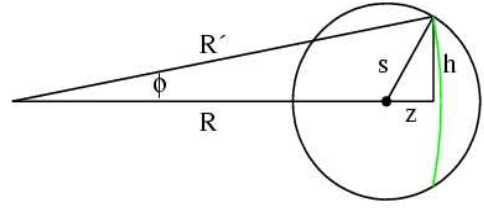,width=3.2in,height=1.8in}
\end{minipage}
\hfill
\begin{minipage}[t]{3.2in}
\psfig{figure=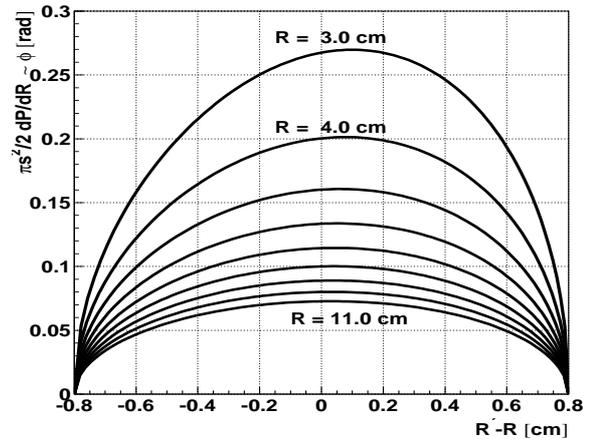,width=3.2 in,height=2.5in}
\end{minipage}
\caption{The intersection of a ring with a tube isn't symmetric.
         The acceptance is biased to R beyond the tube center  [top].
         The acceptance function as a function of ring radius [bottom].}
\label{TUBE}
\end{figure}

%
We've identified several systematic effects.  The first is just geometry;
as shown in Figure~\ref{TUBE}(top), the intersection of a circular ring with 
a circular tube isn't symmetric.  The acceptance correction (proportional the 
the arc length shown) is biased towards radii larger than the center of 
ring to center of tube distance.  This acceptance is a just geometric.  It is 
plotted as a function of $z=R'-R$ in Figure~\ref{TUBE}(bottom) for different 
ring radii.

We've build a second ring radius fitter based on maximizing the joint 
likelihood for all tubes on a ring where the likelihood for each tube is from 
the acceptance curves shown.  The average shift in ring radius is 
$\sim 0.1mm$ or $0.1-0.2\%$ for rings in the radius interval ($5-8$ cm) we use 
to fit the mass.
 
     This likelihood fitter has an interesting application in pattern 
recognition.  The acceptance plotted goes exactly to zero at the radius 
difference where the ring no longer crosses the hit tube.  To deal with these 
cases we assign a noise probability of $0.002$ for the minimum likelihood of 
any tube in the ring fit.  The likelihood as a function of fit radius can have 
one or more maxima.  Consider the case of two tubes along a ring radius: only
one can fit the ring, the other being ``noise'' in this model.  This likelihood
has two maxima (if all the other tubes in the fit are well behaved).  Rejecting
these ambiguous fits is an objective way to remove non-Gaussian tails due 
to pattern recognition mistakes from the ring radius resolution function.

Repeating the mass analysis after applying cuts based on the criteria outlined 
above yield the mas values shown in Figure ~\ref{MASS}.  The 
$\chi^{2}$ for the mass averages as a function of radius are improved but 
systematics in the mass as a function of ring radius remain for the pions and 
kaons when the calibration constants are determined using the protons. 

%
     We have made some progress.  The proton and anti-proton mass now agree 
with each other.  We used the protons for calibration so the proton mass has to
agree with the PDG value.  The kaon masses are close to each other and to the 
PDG values but still several 100 KeV and several $\sigma$ off.  The pion masses
are still quite far from the PDG values for reasons which are unclear.  We can 
and will apply these methods to the charged hyperons when the systematics are 
better understood.  

     This study is beginning to illuminate some of the systematics of the 
resolution of RICHs as precision velocity spectrometers.  More work will be
required to reach and understand the resolution limits of this technique.

%
We are indebted to Selex for the RICH data, our home institutions, 
Consejo Nacional de Ciencia y Tecnolog\'{\i}a {\nobreak (CONACyT), Fondo
de Apoyo a la Investigaci\'on (UASLP), and the U.S. Department of Energy 
(contract DE-AC02-76CHO3000), for support.

\end{document}